\documentclass{iau}
\usepackage{graphicx}
\usepackage{gensymb}
\usepackage{amsmath}
\usepackage{amssymb}

\title[Cloud infall in the Galactic Centre] 
{Forming misaligned stellar discs around a massive black hole: \\Cloud infall in the Galactic Centre}

\author[William Lucas et al.]   
{William Lucas$^{1,*}$, Ian Bonnell$^1$, Melvyn Davies$^2$ \and Ken Rice$^3$}

\affiliation{$^1$ SUPA, School of Physics \& Astronomy, University of
    St Andrews, North Haugh, St Andrews, Fife KY16 9SS, United Kingdom \\
   $^*$ email: wel2@st-andrews.ac.uk \\ 
     $^2$ Lund Observatory, Department of Astronomy and Theoretical Physics, 
    Box 43, SE-221 00 Lund, Sweden \\
     $^3$ SUPA, Institute for Astronomy, University of
    Edinburgh, Blackford Hill, Edinburgh EH9 3HJ, United Kingdom}

\pubyear{2013}
\volume{303}  
\pagerange{1--2}
\jname{The Galactic Center: Feeding and Feedback in a Normal Galactic Nucleus }
\editors{L. Sjouwerman, J. Ott \& C. Lang, eds.}
\begin{document}

\maketitle

\begin{abstract}

The innermost parsec around Sgr A* has been found to play host to two discs or streamers of O and W-R stars. They are misaligned by an angle approaching $90\degree$. That the stars are approximately coeval indicates that they formed in the same event rather than independently. We have performed SPH simulations of the infall of a single prolate cloud towards a massive black hole. As the cloud is disrupted, the large spread in angular momentum can, if conditions allow, lead to the creation of misaligned gas discs. In turn, stars may form within those discs. We are now investigating the origins of these clouds in the Galactic Centre (GC) region.

\end{abstract}

\section{Misaligned discs around a black hole}
\begin{figure}\begin{center}
  \includegraphics[height=3.0in]{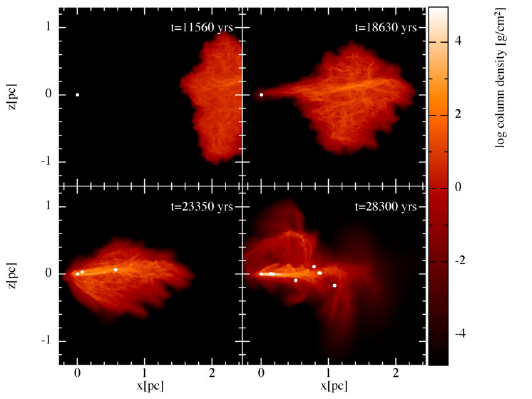} 
 \caption{Here we show in $x$ and $z$ coordinates the column densities in a turbulent cloud of $10^4M_\odot$ falling inwards towards a BH of $4\times10^6 M_\odot$, represented by the filled circle at the origin. It is prolate to the orbital plane (i.e. elongated in the $z$-direction). This increased the spread in the direction of angular momentum within the cloud to the point that the positive and negative $z$ regions orbited the BH in almost opposite directions. The final panel shows that the bottom region survived to form a streamer. Its orbital plane lay about $60 \degree$ from the main disc, which itself lay in the $x$-$y$ plane and can here be seen edge-on. Several stars formed, and are shown as further filled circles. Nine were bound to the BH by the final time. This simulation is presented as Run A in \cite[Lucas et al. (2013)]{Lucas13}; the figure was produced with SPLASH (\cite[Price 2007]{Price07}). }
   \label{fig1}
  \end{center}
\end{figure}

\cite[Paumard et al. (2006)]{Paumard06}, among others, have detected the presence of two misaligned discs of massive stars on orbits within $0.5\,\textrm{pc}$ of Sgr A*. The fragmentation of a gaseous disc around a massive black hole (BH) has been shown to lead to the formation of a corresponding stellar disc, following the infall and tidal shearing of a massive cloud (\cite[Bonnell \& Rice 2008]{BonnellRice08}). With this in mind, we have run a suite of smoothed particle hydrodynamic (SPH) simulations showing the infall of a gas cloud towards a black hole of $4\times 10^6 M_\odot$, representing Sgr A*. These have been presented in \cite[Lucas et al. (2013)]{Lucas13}. In Figure~\ref{fig1} we show the evolution of Run~A from this paper.

In all of our simulations we used a prolate cloud, the major axis oriented perpendicular to the orbital plane. When its initial velocity was close to radial (that is to say, the tangential component was small) this geometry brought the upper and lower regions of the cloud to the point where they rotated in almost opposite directions around the BH. This spread in angular momentum allowed the formation of a misaligned disc-streamer system. The parameter space that allows this outcome is somewhat restrictive, however. Firstly, the orbit must be highly radial in order to generate a large enough difference in the direction of angular momentum within the cloud. Secondly, the cloud must be structured asymmetrically above and below the orbital plane. Otherwise angular momentum will be cancelled out in shocks between gas flowing in opposite directions around the BH, and no streamer will form.
 
\section{Formation of clouds in the inner Galactic Centre}

We are currently investigating whether the effect of tidal forces on gas in the GC may bring about the formation of massive clouds such as G0.253+0.016 (`the Brick'). In SPH we have created a cloud of $10^6 M_\odot$, with radius $17\,\textrm{pc}$ and followed its evolution as it moves through the GC potential. 

Its orbit had pericentre $\sim 15\,\textrm{pc}$ and apocentre $\sim 100 \,\textrm{pc}$. We found that the tidal forces decreased the cross-section of the flow of gas, while also shearing it in the direction of motion. The extension of gas along the non-closing orbit led to the formation of a D-shaped ring with a `corner' formed by the intersection of the leading and trailing gas. This ring, with the central BH offset from its centre, and the asymmetric distribution of mass, is intriguingly similar to Herschel observations of the inner $100\,\textrm{pc}$ of the Galaxy described by \cite[Molinari et al. (2011)]{Molinari11}. We are also interested in how the angular momentum within the gas may be altered at the intersecting point of the ring, particularly as this could potentially lead to infall of gas towards the central BH. We are currently investigating further.

\end{document}